\RequirePackage{amsmath}

\documentclass[12pt,preprint]{iopart}

\usepackage{amsmath}
\usepackage{pdfpages}

\usepackage{graphicx, hyperref, bm, xcolor}
\usepackage[normalem]{ulem}
\usepackage{dcolumn}
\usepackage{acronym}
\graphicspath{ {./pictures/} }
\usepackage{siunitx}
\usepackage[capitalise]{cleveref}
\usepackage{todonotes}

\DeclareSIUnit\samples{samples}

\acrodef{SCF}{sine-cos-fit}
\acrodef{FPD}{free precession decay}
\acrodef{MEOP}{metastability exchange optical pumping}
\acrodef{OPM}{optically pumped magnetometer}
\acrodef{VP}{variable projection}
\acrodef{FFT}{fast Fourier transformation}
\acrodef{SNR}{signal-to-noise ratio}
\acrodef{EKS}{extended Kalman smoother}
\acrodef{EKF}{extended Kalman filter}
\acrodef{NMR}{nuclear magnetic resonance}
\acrodef{CRLB}{Cramér–Rao lower bound}

\begin{document}

\title[EKS of FPD Signals]{Extended Kalman Smoothing of Free Spin Precession Signals for Accurate Magnetic Field Determination}

\author{Jasper Riebesehl$^\dagger$}
\address{Department of Electrical and Photonics Engineering, Technical University of Denmark (DTU), DK-2800 Kgs. Lyngby, Denmark}
\ead{janri@dtu.dk}

\author{Lutz Mertenskötter$^\dagger$}
\address{Weierstrass-Institut fur Angewandte Analysis und Stochastik (WIAS), Mohrenstraße 39, 10117 Berlin, Germany}
\ead{mertenskoetter@wias-berlin.de}

\author{Wiebke Pohlandt}
\address{Physikalisch-Technische Bundesanstalt (PTB), Abbestr. 2-12, 10587 Berlin, Germany}
\ead{wiebke.pohlandt@ptb.de}

\author{Wilhelm Stannat}
\address{Institut für Mathematik, Technische Universität Berlin, Straße des 17. Juni 136, 10623 Berlin, Germany}
\ead{stannat@math.tu-berlin.de}

\author{Wolfgang Kilian}
\address{Physikalisch-Technische Bundesanstalt (PTB), Abbestr. 2-12, 10587 Berlin, Germany}
\ead{wolfgang.kilian@ptb.de}


\vspace{10pt}
\begin{indented}
\item[]$^\dagger$ These authors contributed equally to this work.
\item[]\today 
\end{indented}
\newpage
\begin{abstract}
We present a novel application of the Extended Kalman Smoother (EKS) for highly accurate frequency estimation from free spin precession signals of polarized $^3$He. Traditional approaches often rely on nonlinear least-squares fitting, which can suffer from limited robustness to signal decay and time-dependent frequency shifts. By contrast, our EKS-based method captures both amplitude and frequency variations with minimal tuning, adapting automatically to fluctuations via an expectation-maximization algorithm.

We benchmark the technique in extensive simulations that emulate realistic spin precession signals with exponentially decaying amplitudes and noisy frequency drifts. Compared to least-squares fits with fixed block lengths, EKS systematically reduces estimation errors, particularly when frequencies evolve or signal-to-noise ratios are moderate to high. We further validate these findings with experimental data from a free-precession decay $^3$He magnetometer.

Our results indicate that EKS-based analysis can substantially improve accuracy in nuclear magnetic resonance-based magnetometry, where accurate frequency estimation underpins absolute field determinations. This versatile approach promises to enhance the stability and accuracy of future high-precision measurements.\\
\end{abstract}

\vspace{2pc}

%
\maketitle
%
%


\section{Introduction}

\paragraph{Spin precession magnetometry}
 Magnetometry, i.e. the determination of the static magnetic flux density, is an extremely broad field of research in which a wide variety of technical methods are used \cite{Grosz2017_HSM}. The potentially most precise technique which allows to directly trace back the unit of Tesla to the SI unite of the second is the measurement of the free spin precession frequency using either electronic spin resonance (ESR) or \ac{NMR}. Given the fundamental relation of the Larmor resonance frequency $f_\mathrm{L}$ and magnetic field strength $B_0$
\begin{equation}
    f_{\mathrm{L}} = \frac{\gamma}{2\pi}B_0,
    \label{eq:Larmor-f}
\end{equation}
only the substance-specific gyromagnetic ratio $\gamma$ has to be known to deduce the external magnetic field strength from Larmor precession measurements. 
The gyromagnetic ratio $\gamma_p'$ of protons has been known for decades with a relative uncertainty well below $10^{-6}$ and has more recently been determined with an uncertainty even below $10^{-8}$ \cite{CODATA_2022_online}. 
Consequently, the classical proton free induction decay (FID) measurement -- aside from its widespread use in chemical analysis and medical imaging -- has naturally become a prominent method for highly accurate magnetic field measurements  \cite{Fei1997_NIM-A_349, Weyand2001_IEEE_TIM_470}. With recent advances in measuring the magnetic moment of $^3$He$^+$ and calculation of the diamagnetic shielding, the gyromagnetic ratio $\gamma_h'$ of gaseous $^3$He is now known with even higher accuracy, achieving a relative uncertainty below $10^{-9}$ \cite{Schneider2022_N_878, CODATA_2022_online}. 
Given these precisely known scaling factors for $\gamma'_p$ and $\gamma'_h$, the overall accuracy of magnetic field measurements now critically depends on the precision of the frequency estimation from NMR signals. Especially, the ability to track magnetic field fluctuations over time is a task which the frequency estimator has to fulfill with the utmost accuracy.

In this work we employ a new frequency tracking technique, based on frequency domain Kalman smoothing to improve this accuracy.
Consequently, $^3$He \ac{NMR} provides accurate magnetic field measurements, provided that the systematic errors of the magnetometers
themselves are properly accounted for \cite{Farooq2020_PRL_223001}. These systematic errors are not further discussed in this work, rather we focus on the accuracy of the frequency estimators in terms of trueness, as measured by estimator bias, and precision, as measured by estimator variance over a long measurement period with respect of field drifts\cite{ISO5725-1:2019}.

A wide range of analytical methods -- and their variants -- have been proposed for this task, including frequency-domain analysis using \ac{FFT} \cite{Stamataki2013_AO_1086, Moriat2005_ToneExtraction_patent}, zero-crossing counting \cite{Yurin2024_MNRAS_1483}, singular value decomposition \cite{Lin1997_JMR_30}, Hilbert transform \cite{Hong2021_JMR_107020}, demodulation \cite{Kamwa2004_IEEE-TPD_505514}, and separable nonlinear least-squares analysis using the variable projection method \cite{Gene_Golub_2003}. More recently, machine learning-based approaches have also been explored \cite{Visschers2021_MLST_45024, Almayyali2021_S_2729}.

To our knowledge, the nonlinear least-squares method remains the benchmark for estimating frequencies in ultra-high-precision \ac{NMR} signals in $^3$He-$^{129}$Xe-co-magnetometry measurements \cite{Gemmel2010_EPJD_303, tullney_constraints_2013, Sachdeva2019_PRL_143003}. 
Our frequency domain Kalman smoothing method is a promising candidate for a more accurate estimator, as it effectively filters out the majority of the noise, and -- unlike the least-squares method -- is able to assimilate the data over the entire measurement duration in the presence of magnetic field drift, rather than treating it as a series of uncorrelated blocks. In the following chapters we give an in-depth comparison of this method to the least-squares method on a host of simulated data and real measurements. 

The Kalman smoother \cite{rauch_maximum_1965} not only estimates the hidden state \(\mathbf{x}_k\), in our case frequency $f(t)$ and amplitude $A(t)$, but also computes its covariance matrix \(\mathbf{P}_k\). The recursive update of \(\mathbf{P}_k\) provides a robust measure of the confidence in the state estimates, which is crucial in sensor applications, where the uncertainty is of paramount importance. 

Kalman smoothers have found numerous applications in post-processing of sensor data, geophysical signal reconstruction, and time series forecasting. For instance, in biomedical applications, Kalman smoothers improve electroencephalography (EEG) \cite{tarvainen2024}.
In magnetometry, Kalman filters have been successfully used to significantly improve the resolution of atomic sensors \cite{jimenez-martinez_signal_2018}.

\section{Frequency estimation methods}
\subsection{The Kalman smoother}
The Kalman smoother is a recursive algorithm designed to estimate the state of a dynamical system from noisy observations. It is formulated within a state-space framework, which separates the system dynamics from the measurement process. In the nonlinear case, the system dynamics are described by the \emph{state equation} 
\begin{align}
    \mathbf{x}_k = \phi(\mathbf{x}_{k-1}, t_{k-1}) + \mathbf{w}_k, \label{eq:state_eq}
\end{align}
where $\mathbf{x}_k$ represents the state vector at time $t_k$, and $\phi(\mathbf{x}, t)$ is a (potentially nonlinear) function. The \textit{process noise} $\mathbf{w}_k$ is white noise with covariance $Q$. The measurement process, is in turn described by the \emph{measurement equation}
\begin{align}
\mathbf{y}_k = h(\mathbf{x}_k, t_k) + \mathbf{v}_k,    \label{eq:measurement_eq}
\end{align}
 where $h(\mathbf{x}, t)$  is a nonlinear function mapping the state to the observed measurements $\mathbf{y}_k$, and the \textit{measurement noise} $\mathbf{v}_k$ is white noise with covariance $R$. 

The Kalman Smoother then estimates the hidden state $\mathbf{x}_k$ at time $t_k$ given all $N$ measurements, i.e. it computes $p(\mathbf{x}_k | \mathbf{y}_{1:N})$. It does so, such that it minimizes the mean squared error of the estimated state vs. the true state. Strictly, this applies only when $\phi(\mathbf{x}, t)$ and $h(\mathbf{x}, t)$ are linear. Yet, the \ac{EKS} used here performs very well if the linearizations are as adequate as in our case. The Kalman Filter in turn estimates $\mathbf{x}_k$ at time $t_k$ given only the measurements up to time $t_k$, i.e. it computes
$p(\mathbf{x}_k | \mathbf{y}_{1:k})$, as needed in real-time applications.

\subsection{The model}\label{sec:Kalman_filter_model}
In this work we will take a different approach in modeling the atomic sensor, that strays from the traditional conceptualization in which $\phi(\mathbf{x}, t)$ is typically a discretization of a differential equation -- 'the dynamics' -- and $h(\mathbf{x}, t)$ is a function modeling the relationship of the hidden state to the measured quantity   -- 'the measurement'. Rather we focus on the statistics of the data, namely that \eqref{eq:state_eq} and \eqref{eq:measurement_eq} both have additive white noise to derive a more versatile model.

To extract the Larmor frequency of the spin precession signal, we adapted the approach of La Scala et al.\,\cite{la_scala_extended_1996}, which applies an \ac{EKF} to track a harmonic signal with low \ac{SNR} and slowly varying frequency $ f(t)$ as is typical in spin precession measurements. The main idea is to derive the analytic expression for the Fourier coefficients of the harmonic signal \eqref{eq:harmonic_signal_block} and use this as the measurement equation \eqref{eq:measurement_eq} to compare the model prediction to the Fourier coefficients of the signal, rather than using the time-domain model \eqref{eq:harmonic_signal_block} itself. To this end, the signal model is separated into non-overlapping blocks, each of duration $T_{\mathrm{bl}}=\Delta t N_{bl}$ with $N_{\mathrm{bl}}$ sample points (separated by the sampling period $\Delta t$) in which the frequency $  f_{\mathrm{center}}+\delta f_{k}$ and amplitude $A_k$ are assumed to be constant
\begin{align}
z_{k}(t_{n})&=A_{k}\,\cos\left\{2\pi\left[ f_{\mathrm{center}}+\delta f_{k}\right]t_{n}+\varphi_{k}\right\} +\eta_{k}\left(t_{n}\right),\label{eq:harmonic_signal_block}
\end{align} 
with additive Gaussian noise \( \eta(t) \) of variance \( \sigma_\eta^2 \). $\varphi_{k}=\Sigma_{i=0}^{k-1}2\pi\delta f_{k}T_{bl}+\varphi_{0} $ then is the accumulated phase at the beginning of the $k-$th block and $ f_{\mathrm{center}}$ is the part of the frequency that fits integer periods into the block, i.e. $\text{cos}\left(2\pi( f_{\mathrm{center}}+\delta f_{k})T_{bl}\right)=\text{cos}\left(2\pi\delta f_{k}T_{bl}\right)$. The discrete Fourier transform of each block then reads
\begin{align}
\mathcal{F}z_{k}( f_{m})&=\sum_{n=0}^{N_{bl}-1}z_{k}(t_{n})e^{-i 2\pi f_{m}t_{n}}=h_m\left(\mathbf{x_k}\right)+v_{m,k}\\
&=\frac{A_{k}}{2}\left(e^{i\varphi_{k}}\frac{1-e^{i2\pi\delta f_{k}T_{bl}}}{1-e^{i2\pi( f_{\mathrm{center}}+\delta f_{k}- f_{m})\Delta t}}+e^{-i\varphi_{k}}\frac{1-e^{-i2\pi\delta f_{k}T_{bl}}}{1-e^{-i2\pi( f_{\mathrm{center}}+\delta f_{k}+ f_{m})\Delta t}}\right)+v_{m,k},\label{eq:fourier_coeff}
\end{align}
where $, f_{m}=\frac{m}{T_{bl}},m=1,...,N_\mathrm{bl}$.
Since the Fourier transform is linear, the Gaussian white noise \( \eta_k(t) \) transforms into independent Gaussian noise components $v_{m,k}=\mathcal{F}\eta_k\left( f_{m}\right)$ in the Fourier coefficients, with variance of the independent real and imaginary parts
\begin{align}
\sigma_{v}^2 = \frac{2}{N_\mathrm{bl}} \sigma_\eta^2
\end{align}
that is significantly reduced compared to that of $\eta_k$.
Thus, the $\mathcal{F}z_{k}( f_{m})$ themselves form a time series with white Gaussian noise components, and are therefore compatible with the \ac{EKS} framework.

\begin{figure}
    \centering
    \includegraphics[width=.9\linewidth]{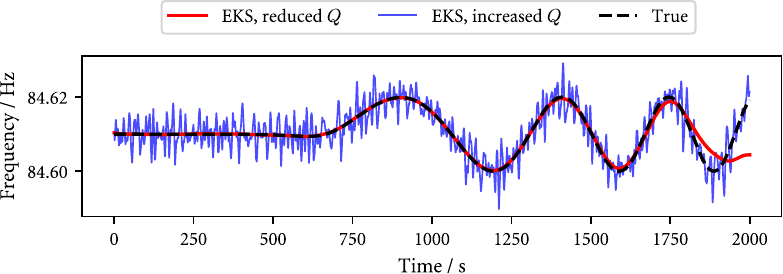}
    \caption{\ac{EKS} tracking of toy example with time-dependent frequency variation chosen to illustrate the variance/stiffness trade-off. The true frequency (black dashed line) is followed by a model with artificially large $Q = 200\, Q_{opt}$ (blue) even when the frequency changes rapidly in time. Meanwhile, a model with artificially small $Q= \frac{1}{200}\, Q_{opt}$ (red) exhibits a maximum slew rate that is at some point exceeded by the rate of change of the frequency. Conversely, where the frequency is constant, the model with small $Q$ has much smaller variance and outperforms the model with large $Q$. Here $Q_{opt}$ is the optimal covariance as determined by the algorithm detailed in the supplementary material.}
    \label{fig:cartoon_chirped_sin}
\end{figure}

The evolution of the hidden variables \eqref{eq:state_eq} is modeled by a typical augmented Kalman filter approach,
\begin{align}
\mathbf{x}_{k} =
\begin{bmatrix}
 A_{k} \\
 \Delta A_{k} \\ 
 \varphi_{k} \\ 
 \delta f_{k}\\
 \Delta \delta f_{k}\\
\end{bmatrix} = \begin{bmatrix}
 A_{k-1} + \Delta A_{k-1}  \\
 \Delta A_{k-1} \\ 
\varphi_{k-1} + 2\pi\delta f_{k-1}  \\ 
 \delta f_{k-1}+ \Delta\delta f_{k-1}  \\
 \Delta\delta f_{k-1}
\end{bmatrix} + \mathbf{w}_k,
\end{align}
with independent integrated random walks in the amplitude and frequency offset, while the phase is only an auxiliary variable that is computed as the numerical integral of the frequency. The process noise $\mathbf{w}_k$ is white noise with covariance 
\begin{align}
Q = \text{diag}\left(\begin{bmatrix}{}
0 & Q_{\Delta A} & 0 & 0 & Q_{\Delta\delta f}\\
\end{bmatrix}\right) \label{eq:Q_model},
\end{align}
such that the variables $A$, $\delta_f$, and $\varphi$ that feed into the model through \eqref{eq:fourier_coeff} are integrated and twice-integrated white noise, respectively. This choice of integrated model for the state evolution reduces the variance of the tracking, but also reduces its slew rate, as the integration amplifies slow components of the model while suppressing fast ones. 
The filter then compares the estimated state not to the signal directly in the time-domain, but rather to its Fourier coefficients \eqref{eq:fourier_coeff}, which serve as the measurement model \eqref{eq:measurement_eq} via

\begin{align}
    \mathbf{y}_k\left(\mathbf{x}_k\right)=\begin{bmatrix}
 h_{M-L}(\mathbf{x}_k)\\
\vdots \\ 
 h_{M+L}(\mathbf{x}_k)\\
\end{bmatrix}+\begin{bmatrix}
 v_{M-L,k}\\
\vdots \\ 
 v_{M+L,k}\\
\end{bmatrix}.
\end{align}

Rather than operating on all Fourier coefficients, the \ac{EKS} only incorporates a range of frequencies $f_{M-L}...f_{M+L}$ around the central frequency $f_M =  f_{\mathrm{center}}$, controlled by the parameter \( L \), which determines how many Fourier bins contribute to the estimation. A larger \( L \) improves robustness to frequency drifts but introduces more noise and thus reduces accuracy, while a smaller \( L \) improves accuracy but reduces adaptability. By tuning \( L \), this method enables reliable frequency tracking even in extreme noise conditions. Frequency drifts in DC magnetometry systems are typically very small, and it is sufficient to choose $L = 1$, such that $2L+1=3$ Fourier bins are used.

 The static parameters of the model
\begin{align}
    \theta=\left\{ Q,R,\delta f_{0}, A_0 \right\}, \label{eq:parameters}
\end{align} that is, the process noise covariance $Q$, the measurement noise covariance $R$, and the starting values $\delta f_0$ and $A_0$ are all optimized separately using an algorithm based on expectation maximization. Details on the algorithm can be found in the supplementary material. An important strength of this model is that \eqref{eq:measurement_eq} and \eqref{eq:state_eq} are entirely free of system-dependent parameters that would need to be known in order to perform the tracking. The only static parameters in the Kalman smoother itself are noise-covariances and starting values in \eqref{eq:parameters}, which are inherently present in any state-space filtering problem. Optimization of the process covariances $Q$ is particularly vital to the performance of the \ac{EKS}, as the tracking becomes too stiff when it is too small, while a $Q$ that is too large results in a large variance of the estimates, as is illustrated in a toy example in \cref{fig:cartoon_chirped_sin}.

\subsection{Least Square Analysis as Reference Method}
To separate the nonlinear signal dependence on frequency $f$ and phase offset $\varphi_0$ at $t=0$ for a signal obeying $S(t)=A \sin{\left( 2\pi f t+\varphi_0\right)}$ within the least squares framework, an equivalent model function, commonly referred to as the \ac{SCF} is used
\begin{equation}
    S(t) =  A_s \sin\left(2\pi f t\right) + A_c\cos\left(2\pi f t\right) + C_0 ,
\label{eq:sincosfit_new}
\end{equation}
which allows to calculate the signal amplitude of the corresponding data set by $A=\sqrt{A_s^2+A_c^2}$ and the phase at $t=0$ by $\varphi_0=\arctan{\left( A_c / A_s \right)}$ and accounts for a constant signal offset $C_0$.
%
%
%
Fitting the data to  \eqref{eq:sincosfit_new} permits the use of the \ac{VP} method \cite{Gene_Golub_2003} wherein the linear terms are estimated separately from the nonlinear frequency dependence. The method alternates between a linear least-squares fit of the linear parameters $A_s$, $A_c$, and $C_0$, and a non-linear least-squares fit of the frequency $f$ using the Levenberg-Marquardt algorithm. The uncertainties of the four parameters $u_{f}$, $u_{Ac}$, $u_{As}$, and $u_{C0}$ are obtained by scaling the covariance matrix with the mean squared errors of the residuals. The covariance matrices for $f$ and the other three parameters are obtained from the two separate fitting procedures of the \ac{VP}, respectively.
The start parameter for $f$ for the fitting process, is determined beforehand by performing a \ac{FFT} on the complete dataset and searching for the frequency bin with the highest peak in a given range. The start values for the parameters $A_s$, $A_c$, and $C_0$ are set to 1, since their estimation via linear least squares is largely insensitive to the choice of starting values.

As the function \eqref{eq:sincosfit_new} does not take into account the exponential decay of the amplitude, the entire time domain data is divided into non-overlapping blocks of equal length. These blocks are chosen to be short enough that the signal amplitude can be assumed constant, yet long enough to sufficiently reduce the variance of the least-squares estimates. As the uncertainty (i.e., standard deviation) of the fitted parameters scales inversely with the square root of the number of sample points, assuming independent, homoscedastic Gaussian noise \cite{Kay1993_Statistical_Signal_Processing}, in principle a longer block length should reduce the uncertainties. However, in our experimental data this is compromised by the signal amplitude decay and frequency drifts due to a drift in $B_0$. 

\section{Comparing the methods on simulated data}
To quantitatively determine which of the methods can provide better frequency estimates in \ac{FPD} signals, we turn to a broad simulation study. The characteristic properties of such a signal are its exponential amplitude decay and the time-dependence of the true frequency, which is in turn determined by the fluctuations of the true magnetic field strength.
To represent this class of signals, we use the phenomenological model
\begin{align}
    y(t_i) = A_0 \, e^{-t_i/T_2^*} \, \sin \left\{2\pi f(t_i)t_i + \varphi_0 \right\} + \eta(t_i),
    \label{eq:sim_signal}
\end{align}
where $A_0$ is the initial signal amplitude, $T_2^*$ the effective time constant of the transverse spin-relaxation leading to loss of phase-coherence and thus to exponential signal decay, $f(t)$ the time-varying frequency, $\varphi_0$ an initial phase and finally, $\eta(t)$ subsumes all noise sources accumulated within the measurement data. The index $i = 0 \dots N$ indicates a discretely sampled signal with sampling rate $f_s$.

We generated simulated signals for different value combinations of these parameters to evaluate the methods on. We are mainly interested in investigating the influence of $A_0$ and the strength of frequency fluctuations, while $T_2^*$ and the variance of $\eta$ are fixed for a certain experimental setup. 

The stochastic nature of field fluctuations is emulated by modeling the time dependence of the frequency with a random walk plus a constant offset $f_c$. An instance of a random walk is generated by cumulatively summing discrete random increments $\delta f$ drawn from a Gaussian distribution with variance $\sigma^2_{\delta f}$.
The magnitude of the frequency drift is quantified by the diffusion constant $D$ of the random walk, with
\begin{align}
    D = \frac{\sigma^2_{\delta f} f_s}{2}.
\end{align}
The second parameter to be varied is the initial amplitude $A_0$, giving an initial \ac{SNR}
\begin{align}
    \mathrm{SNR}_0 = \frac{A_0^2}{2\sigma_{\eta}^2}.
\end{align}
For simplicity, $\eta(t)$ is white Gaussian noise with fixed variance $\sigma_{\eta}^2$. Real experimental signals, however, can also contain low-frequency magnetic field perturbations and additional noise signals with frequency components that do not originate from the spin precession, e.g. from power-line or setup vibrations. Because the Kalman smoother method effectively applies a band-pass filter these deviations from the model can be neglected unless they are close in frequency to the signal itself.

Due to the stochastic nature of the simulated signals, multiple repetitions were simulated per set of parameters and the ensemble means are reported. The fixed parameters of this study are shown in the supplementary material. They were selected to closely resemble those of the measured data. 
We generated an ensemble of signals according to \eqref{eq:sim_signal} for a range of values of $D$ and $\mathrm{SNR}_0$. For each instance, both the \ac{SCF} and the \ac{EKS} methods were applied.

\begin{figure}[h]
    \centering
    \includegraphics[width=0.9\linewidth]{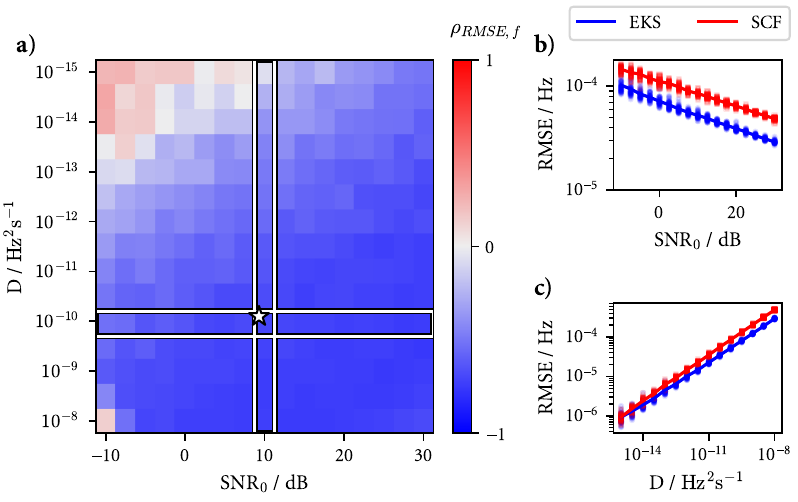}
    \caption{Quantitative comparison of the \ac{SCF} against the \ac{EKS} in a simulation study. \textbf{a)} $\log_2$ of the mean ratio of the frequency estimate errors. Blue indicates a better performance of the \ac{EKS} over the \ac{SCF}, red the opposite. The star marker indicates a rough estimate of our experimental conditions. \textbf{b)} Horizontal slice through a). The RMSEs of individual simulation runs are displayed as markers, the solid lines indicate the mean. \textbf{c)} displays a vertical slice.}
    \label{fig:sim_f_heatmap}
\end{figure}

For the \ac{SCF}, a broad range of block sizes were evaluated for each $D$ and SNR$_0$ independently. Ultimately, the block size which results in the estimate with the lowest MSE was selected as the best estimate (supplementary material\,2.4). Note that this is the best case for the \ac{SCF} estimate, and this selection method is only available in simulation.

The application of the \ac{EKS} includes the full parameter optimization routine using the EM algorithm. Here, we intentionally kept the initial parameters for the optimization fixed to emulate zero prior knowledge about the signal. This demonstrates the method's robustness as the optimization's convergence does not critically depend on the initial parameters. The only parameter which has to be adapted by simulation studies is $T_{bl, \mathrm{EKS}}=4.5\,$s providing valid uncertainty bounds for the \ac{EKS} (supplementary material\,2.3).

\subsection{Results}
The results of this study are summarized in \cref{fig:sim_f_heatmap}. The frequency estimation errors of both methods are compared by calculating the ratio of their root mean squared errors 
\begin{align}
    \rho_{\mathrm{RMSE}, f} = \log_2 \frac{\mathrm{RMSE}_{f, \mathrm{EKS}}}{\mathrm{RMSE}_{f, \mathrm{SCF}}},
\end{align}where
\begin{align}
    \mathrm{RMSE}_f= \sqrt{ \frac{1}{M}\sum_{j=1}^M\frac{1}{N}\sum_{i=1}^N\left(f_{j}^\mathrm{est}(t_i)-f_{j}^\mathrm{true}(t_i)\right)^2}
    \end{align} 
is the root average of the square deviation of the estimated value $x^\mathrm{est}$ from the true value $x^\mathrm{true}$ over all $N$ times and $M$ realizations. Because the estimates of the EKS and SCF have different block lengths we up-sample both to the resolution of the ground truth to achieve a fair comparison.
The up-sampling extrapolates the estimates per block to constant values across all samples per block.
$\rho_{\mathrm{RMSE}, f}$ is zero when the errors are equal, becomes negative for lower error of the \ac{EKS} frequency estimate, and positive for the opposite case. For a large portion of the parameter space in \cref{fig:sim_f_heatmap}\,a), $\rho_{\mathrm{RMSE}, f}$ is close to \num{-1}, indicating an improved frequency estimate of the \ac{EKS} by a factor of \num{2}. For low SNR and very low frequency drift, $\rho_{\mathrm{RMSE}, f}$ becomes positive, indicating that the \ac{SCF} estimate should be used in these regions.
This seems consistent with the observation that the \ac{SCF} operates closer to the \ac{CRLB} for no frequency drift (see supplementary material\,2.5). Continuously reducing the frequency drift asymptotically approaches this regime. In all regimes the inaccuracy, as measured by RMSE is dominated by the contribution of the estimator variance, while the contribution of estimator bias is consistently orders of magnitude smaller (see supplementary material\,2.6)

We also compared the amplitude estimates, where the \ac{EKS} significantly outperforms the \ac{SCF}.
This is to be expected, as the models don't consider the exponential signal decay in order to stay free of unknown parameters. For the longer \ac{SCF} block lengths, this model mismatch leads to a stronger degradation of estimation performance in the time-dependent amplitude (see supplementary material\,2.2).
\newpage
\section{Application to experimental data}
\begin{figure}[b]
        \centering
        \includegraphics[width=.9\linewidth]{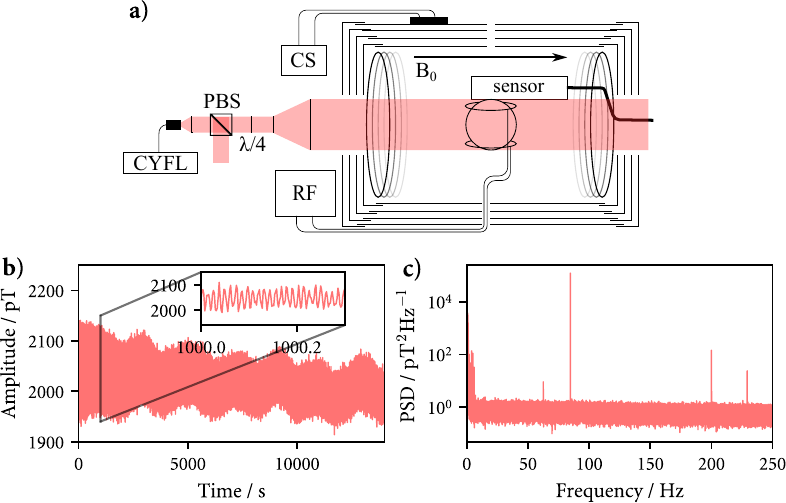}
        \caption{\textbf{a)} Schematic of the setup as used for $^3$He spin precession measurements. For \ac{MEOP}, light from a laser (CYFL) via some optics (PBS and a $\lambda$/4) is shone on the metastable $^3$He generated by a radio-frequency (RF) discharge with electrodes around the spherical gas cell. A current source (CS) attached to a coil inside the four-layer shield generates the $B_0$ field. The dual-cell \ac{OPM} gradiometer sensor is used for reading out the $^3$He spin precession signal. \textbf{b)/c)} Experimental gradiometer signal from the dual-cell \ac{OPM} sensor. \textbf{b)} Time domain of the signal. The inset shows the fast oscillations caused by the $^3$He spin precession. \textbf{c)} Power spectral density of the signal showing the dominant $\approx\,$85\,Hz $^3$He signal.}
        \label{fig:exp_signal}
\end{figure}

Finally, we assess both methods on real experimental data. For this $^3$He spin precession measurements were performed by a setup as sketched in \cref{fig:exp_signal}\,a). A glass cell of 3\,cm diameter filled with 10\,torr $^3$He was positioned within a mu-metal shield which allows to generate a constant holding field $B_0$ in the \SI{}{\uT}-range and a pulsed $B_1$-field ($\perp B_0$) to induce the $\pi/2$ spin flip. Signal detection of the \ac{FPD} was performed alike shown in \cite{Koch2015_EPJD_202}, applying a commercial dual-cell \ac{OPM} in gradiometer arrangement. To maintain sufficiently high signal \ac{MEOP} technique \cite{2017Gentile_R} was used to polarize nuclear spins of the $^3$He atoms. 

Due to the $T_2^\star$ time of about 50\, minutes, signals over a time-span of nearly four hours could be measured as shown in \cref{fig:exp_signal}\,b). The inset shows a zoomed-in view of the sinusoidal signal stemming from the precessing $^3$He magnetization which is generating a time-varying field difference seen by the two Rb cells. The $\approx\,50\,$pT amplitude can be calculated to stem from $\approx\,8$\% $^3$He polarization \cite{Koch2015_EPJD_262}. The large offset of $\approx\,2\,$nT reflects the background field gradient of $\approx\,$\SI{1}{\nano\tesla/\cm} explaining the relatively short $T_2^\star$ time as compared to measurements obtained in a large shielded room \cite{Gemmel2010_EPJD_303} and common in small shielding environment \cite{Klinger2023_PRA_44092}. 
The dominant peak at $\approx\,$\SI{84.6}{\hertz} in the power spectrum in \cref{fig:exp_signal}c) stems from the \ac{FPD} signal corresponding to a constant background field of $\approx\,$\SI{2.61}{\uT}.

\begin{figure}[b]
    \centering
    \includegraphics[width=0.9\linewidth]{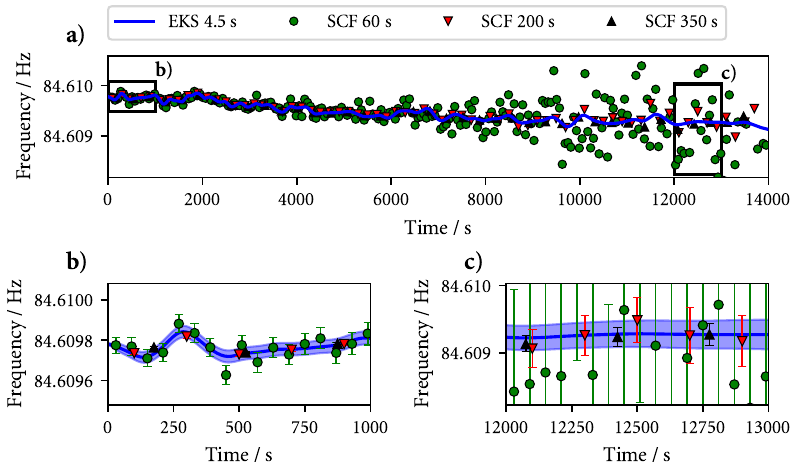}
    \caption{Frequency tracking on experimental data, comparison of the methods. Uncertainty intervals are two standard deviations. \textbf{b)} and \textbf{c)} are zoomed insets of \textbf{a)}.}
    \label{fig:exp_results}
\end{figure}

Both analysis methods were applied directly to the dual-cell \ac{OPM} gradiometer time-domain signal without any pre-processing. Again, $T_{bl, \mathrm{EKS}}=4.5\,$s was used for the \ac{EKS} as determined in the simulation study.
For the \ac{SCF}, multiple block lengths were evaluated as no clear criterion is available to select the optimal block length in the RMSE sense purely from data. The simulation study indicated that the optimal block length for a signal with the given parameters should be close to \SI{200}{\second} (supplementary material\,2.4).

The frequency estimates over time are shown in \cref{fig:exp_results}. In a), the estimates over the full time series are shown without error bars for visibility. The \ac{SCF} estimates of any block lengths are scattered around the \ac{EKS} estimate with increasing variance at larger times. This is caused by the exponential amplitude decay and subsequent decrease in \ac{SNR}.
\cref{fig:exp_results}\,b) shows a zoom to the beginning of the measurement, where the \ac{SNR} is high. Here, the effect of a reduced time resolution when selecting very large block lengths for the \ac{SCF} becomes apparent. For the smaller block length of 60\,s the tracking of fluctuation dynamics is possible, however, resulting in higher uncertainty margins, a typical bias-variance tradeoff. Note that the \ac{SCF} with the largest block length applied here can not resolve the upwards swerve at $\approx\,$\SI{300}{s} while the uncertainty interval is much smaller than this variation, clearly seen with the shorter block lengths and the \ac{EKS}. This hints at unreliable uncertainty estimates of the \ac{SCF} for too large block lengths, however, with no analytical measure to judge on this.

In \cref{fig:exp_results}\,c) the frequency estimates towards the end of the measurement with low SNR are shown. All results agree within their uncertainty intervals. Notably, the \ac{EKS} uncertainty have a similar magnitude to \ac{SCF} uncertainty intervals which use close to 2 orders of magnitude larger block length. An obvious improvement to the \ac{SCF} method would be the introduction of an \ac{SNR}-adaptive block length in future work. However, it is not clear which criterion should be used to determine said block length. 

These results highlight the flexibility of the \ac{EKS}, as it does not need to be tuned manually to perform best in a wide range of parameter regimes. All tuning is handled automatically by the EM algorithm.
 
\section{Conclusion}
In this work we have demonstrated the use of extended Kalman smoothing for accurate time-dependent frequency estimation in $^3$He nuclear spin precession signals aimed to be used for absolute field tracking. Through simulation studies, we have shown that the \ac{EKS} outperforms the nonlinear least squares fit based state of the art method for a wide range of experimental conditions. In addition, we have demonstrated the robustness of the method by applying it to an experimental signal with significant perturbations.
For accessibility, we have provided a robust and easy-to-use implementation of the method.

\section*{Data availability statement}
The code that supports the findings of this study is available online \cite{Riebesehl_KalmanMagnetometry2025}.

\section*{Acknowledgment}
This work has been funded by the SPOC Center (Grant No. DNRF 123), and the German Research Foundation (Deutsche Forschungsgemeinschaft, DFG) under Germany's Excellence Strategy \textendash{} EXC2046: MATH+ (project AA2-13).
The authors thank Silvia Knappe-Grüneberg for valuable discussions and insights that contributed to this work.

\section*{References}
\bibliographystyle{elsarticle-num} 
\bibliography{main}


\section*{Supplementary material}

# Supplementary Material: Extended Kalman Smoothing of Free Spin Precession Signals for Precise Magnetic Field Determination

## 1. Mathematical details on the extended Kalman smoother method

In this section, the equations for the Kalman smoothing approach presented in the main text are given for reproducibility. For derivations and more detail please refer to [1].

### 1.1. Extended Kalman filtering and smoothing equations

The EKS discussed in this work relies on a forward-backward recursion over the signal to be applied. In the first forward recursion, the extended Kalman filter is applied by using the filtering equations given as

$$
\begin{aligned}
\hat{\mathbf{x}}_k &= \phi(\mathbf{x}_{k-1}, t_k) \\
\hat{\mathbf{P}}_k &= J_\phi(\mathbf{x})\Big|_{\mathbf{x}_{k-1}} \mathbf{P}_{k-1} J_\phi(\mathbf{x})^T\Big|_{\mathbf{x}_{k-1}} + \mathbf{Q}_k \\
\mathbf{S}_k &= J_h(\mathbf{x})\Big|_{\hat{\mathbf{x}}_k} \hat{\mathbf{P}}_k J_h(\mathbf{x})^T\Big|_{\hat{\mathbf{x}}_k} + \mathbf{R}_k \\
\mathbf{K}_k &= \hat{\mathbf{P}}_k J_h(\mathbf{x})^T\Big|_{\hat{\mathbf{x}}_k} \mathbf{S}_k^{-1} \\
\mathbf{v}_k &= \mathbf{y}_k - h(\hat{\mathbf{x}}_k) \\
\mathbf{x}_k &= \hat{\mathbf{x}}_k + \mathbf{K}_k \mathbf{v}_k \\
\mathbf{P}_k &= \hat{\mathbf{P}}_k - \mathbf{K}_k \mathbf{S}_k \mathbf{K}_k^T
\end{aligned}
\tag{1}
$$

where $J_\phi(\mathbf{x})\Big|_{\mathbf{x}_k}$ indicates the Jacobian of a function $\phi$ with respect to $\mathbf{x}$, evaluated at $\mathbf{x}_k$. For $k = 0$ an initial state $\mathbf{x}_0$ and its covariance $\mathbf{P}_0$ are required. In our method, we keep $\mathbf{Q}_k$ and $\mathbf{R}_k$ constant over $k$.

In the second step, the EKS estimates are calculated using the smoothing equations:

$$
\begin{aligned}
\mathbf{G}_k &= \mathbf{P}_k \left[ J_\phi(\mathbf{x})\Big|_{\mathbf{x}_k} \right]^T \hat{\mathbf{P}}_{k+1}^{-1} \\
\mathbf{x}_k^s &= \mathbf{x}_k + \mathbf{G}_k \left[ \mathbf{x}_{k+1}^s - \hat{\mathbf{x}}_{k+1} \right] \\
\mathbf{P}_k^s &= \mathbf{P}_k + \mathbf{G}_k \left[ \mathbf{P}_{k+1}^s - \hat{\mathbf{P}}_{k+1} \right] \mathbf{G}_k^T
\end{aligned}
\tag{2}
$$

This backwards recursion is initialized at $k = N$ with $\mathbf{x}_N^s = \mathbf{x}_N$ and $\mathbf{P}_N^s = \mathbf{P}_N$. The smoothing estimates $\mathbf{x}_k^s$ for $k = N \ldots 0$ contain the EKS frequency and amplitude estimates. The diagonal elements of the covariance matrices $\mathbf{P}_k^s$ contain the marginal uncertainties of the estimates.

## 1.2. Robust and fast parameter optimization with the Expectation-Maximization algorithm

The filtering and smoothing equations contain static parameters which need to be optimized for the method to perform optimally. These parameters are the process/measurement noise covariance matrices $\mathbf{Q}$ and $\mathbf{R}$ and the initial state and state covariance, $\mathbf{x}_0$ and $\mathbf{P}_0$.

A well documented algorithm for this purpose is the Expectation-Maximization (EM) algorithm [1, 2]. It iteratively maximizes the likelihood function and can be approximately solved using closed form analytic expressions. Using the auxiliary variables

$$\mathbf{C}_k = \mathbf{P}_k^s \mathbf{G}_{k-1}^T + \mathbf{x}_k^s \left[\mathbf{x}_{k-1}^s\right]^T$$

$$\mathbf{\Sigma}_k = \mathbf{P}_k^s + \mathbf{x}_k^s \left[\mathbf{x}_k^s\right]^T$$

$$\mathbf{\Lambda}^{(j+1)} = \frac{1}{T}\sum_{k=1}^{N} \mathbf{\Sigma}_k - \mathbf{C}_k \left[J_\phi(\mathbf{x})\Big|_{\mathbf{x}_k^s}\right]^T - J_\phi(\mathbf{x})\Big|_{\mathbf{x}_k^s} \mathbf{C}_k^T - J_\phi(\mathbf{x})\Big|_{\mathbf{x}_k^s} \mathbf{\Sigma}_{k-1} \left[J_\phi(\mathbf{x})\Big|_{\mathbf{x}_k^s}\right]^T \quad (3)$$

$$\mathbf{\Omega}^{(j+1)} = \frac{1}{T}\sum_{k=1}^{N} \mathbf{v}_k \mathbf{v}_k^T - J_h(\mathbf{x})\Big|_{\mathbf{x}_k^s} \mathbf{P}_k^s \left[J_h(\mathbf{x})\Big|_{\mathbf{x}_k^s}\right]^T$$

the EM update equations are expressed as

$$\begin{aligned}
\mathbf{R}^{(j+1)} &= (1-\alpha_{\mathbf{R}})\mathbf{\Omega}^{(j+1)} + \alpha_{\mathbf{R}}\mathbf{R}^{(j)} \\
\mathbf{Q}^{(j+1)} &= (1-\alpha_{\mathbf{Q}})\mathbf{\Lambda}^{(j+1)} + \alpha_{\mathbf{Q}}\mathbf{Q}^{(j)} \\
\mathbf{P}_0^{(j+1)} &= (1-\alpha_{\mathbf{P}_0})\left[\mathbf{P}_0^s + (\mathbf{x}_0^s - \mathbf{x}_0)(\mathbf{x}_0^s - \mathbf{x}_0)^T\right] + \alpha_{\mathbf{P}_0}\mathbf{P}_0^{(j)} \\
\mathbf{x}_0^{(j+1)} &= (1-\alpha_{\mathbf{x}_0})\mathbf{x}_0^s + \alpha_{\mathbf{x}_0}\mathbf{x}_0^{(j)}.
\end{aligned} \quad (4)$$

Here we have added exponential smoothing to the update equations in the form of a convex combination of the new and previous parameter estimate. In practice, this improves the convergence of all parameters since it allows individual tuning of the convergence speeds. Throughout this work, we have used $\alpha_{\mathbf{R}} = \alpha_{\mathbf{x}_0} = \alpha_{\mathbf{P}_0} = 0.8$ and $\alpha_{\mathbf{Q}} = 0$. $\mathbf{Q}$ is the slowest parameter to converge and this adjustment can help to equalize the parameter convergence speeds.

Due to numerical inaccuracies, it is possible that the covariance matrices loose the positive semi-definiteness property they are required to have. To ensure that $\mathbf{Q}^{(j+1)}$, $\mathbf{R}^{(j+1)}$ and $\mathbf{P}_0^{(j+1)}$ are positive semi-definite, we enforce this property by applying to function

$$\mathbf{A} \rightarrow \mathrm{Re}\left(\sqrt{\mathbf{A}\mathbf{A}^T}\right). \quad (5)$$

While other projections exists, this particular one has proved to be the most reliable.

## 1.3. Bisection-like algorithm for EM

Simply iterating over the EM update equations (4) does eventually lead to convergence. However, it often requires a set of well-selected initial parameters. In addition,

convergence can be slow as the optimum is approached, especially if the initial conditions are not well conditioned. To improve these the convergence speed and to reduce dependence on initial parameters, we add an optional bisection inspired algorithm on top of EM. The algorithm is defined in Alg. 1 in pseudo-code.

---

**Algorithm 1:** Bisection-like algorithm for EM to improve convergence speed of $\mathbf{Q}$.

---

    **Input** : $(\boldsymbol{\beta}, \boldsymbol{\gamma}, \beta_{end}, \Xi, J, J_{init}, J_{max})$

**1** perform $J_{init}$ EM iterations using Eqs. (4);

**2** $j \leftarrow J_{init}$;

**3** **while** *any($\beta_i > \beta_{end}$) and ($j < J_{max}$)* **do**

**4**      perform $J$ EM iterations using Eqs. (4);

**5**      $j \leftarrow j + J$;

**6**      $\mathbf{B} \leftarrow \mathbb{1}$;

**7**      **for** $i = 1..dims(\mathbf{Q})$ **do**

**8**          $\Delta\mathbf{Q} \leftarrow \mathbf{Q}_{i,i}^{(j)} - \mathbf{Q}_{i,i}^{(j-J)}$ ;

**9**          **if** $\Delta\mathbf{Q} > 0$ **then**

**10**              **if** $\gamma_i \geq 0$ **then**

**11**                  $\mathbf{B}_{i,i} \leftarrow \sqrt{\beta_i}$ ;

**12**              **else**

**13**                  $\beta_i \leftarrow \beta_i^{\Xi}$;

**14**                  $\mathbf{B}_{i,i} \leftarrow \sqrt{\beta_i}$ ;

**15**              **end**

**16**              $\gamma_i \leftarrow 1$;

**17**          **else**

**18**              **if** $\gamma_i \geq 0$ **then**

**19**                  $\mathbf{B}_{i,i} \leftarrow \sqrt{1/\beta_i}$ ;

**20**              **else**

**21**                  $\beta_i \leftarrow \beta_i^{\Xi}$;

**22**                  $\mathbf{B}_{i,i} \leftarrow \sqrt{1/\beta_i}$ ;

**23**              **end**

**24**              $\gamma_i \leftarrow -1$;

**25**          **end**

**26**      **end**

**27**      $\mathbf{Q}^{(j)} \leftarrow \mathbf{B}\mathbf{Q}^{(j)}\mathbf{B}^T$

**28** **end**

---

This algorithm applies $J$ EM iterations and amplifies the convergence trend of the diagonal elements of $\mathbf{Q}$. If the matrix element $\mathbf{Q}_{i,i}$ has converged to higher values within $J$ steps, it is multiplied with a factor $\sqrt{\beta_i}$ larger than 1. For a downwards trend, it is multiplied with $1/\sqrt{\beta_i} < 1$. $\beta_i$ is initially large to cause large jumps in $\mathbf{Q}_{i,i}$. Initial

large jumps ensure that a large portion of the parameter space is explored during the optimization. While $\mathbf{Q}_{i,i}$ is closing in on the true value through continuous application of the EM algorithm, $\beta_i$ is shrunk closer to unity using $\beta_i \leftarrow \beta_i^\Xi$ with $0 < \Xi < 1$ whenever the sign of the previous trend amplification had the opposite sign $\gamma_i$. This causes smaller and smaller jumps to home in on the true value for $\mathbf{Q}_{i,i}$. This ensures that $\beta_i$ monotonically shrinks. The process terminates when $\beta_i < \beta_{end}$ with $\beta_{end}$ close to 1.

## 1.4. Initial values for EM parameter optimization

The modifications and tuning of the EM algorithm presented above lead to apparent independence of the convergence of the initial parameters. Hence we are able to use the same hyper- and initial parameters for all applications of the EKS in this work. The only exception are the initial values for amplitude ($A_0^{est}$) and frequency ($\delta_f^{est}$) of the signal, which appear in $\mathbf{x}_0$. They are estimated using the same approach as for the nonlinear least squares method. All other initial parameters are:

**Table 1.** Hyper- and initial parameters for EKS parameter optimization

| Parameter | Value |
|:---:|:---:|
| $Q^{(j=0)}$ | $\mathrm{diag}(10^{-5}, 10^{-5}, 10^{-5}, 10^{-5}, 10^{-4})$ |
| $R^{(j=0)}$ | $\mathrm{diag}(10^2)$ |
| $x_0^{(j=0)}$ | $\left[ \sqrt{A_0^{est}/(4\pi)}, 1.0, \delta_f^{est}, 0, 0 \right]^T$ |
| $P_0^{(j=0)}$ | $\mathrm{diag}(10^{-1}, 10^{-1}, 10^{-1}, 10^{-2}, 10^{-2})$ |
| $\Xi$ | 0.75 |
| $\gamma_i$ | 0 |
| $\beta_i$ | 100 |
| $\beta_{end}$ | $[\beta_i]^{-2^6}$ |
| $J$ | 20 |
| $J_{init}$ | 200 |
| $J_{end}$ | $100 \cdot J$ |

## 2. Further simulation study results

Here we present further results from the simulation study shown in the main text, but were left out due to space constraints.

## 2.1. Simulation Study Parameters

**Table 2.** Fixed parameters in the simulation study.

| Parameter | Value |
|:---:|:---:|
| N | 5.4 Msamples |
| $f_s$ | 500 Hz |
| $f_c$ | 84.06 Hz |
| $T_2^*$ | 3142 second |
| $\sigma_\eta$ | 10 pT |
| $\varphi_0$ | 0 |
| $N_{repetitions}$ | 50 |
| $N_{bl,EKS}$ | $4.5\,\text{s} \cdot f_s$ |

## 2.2. Amplitude RMSE comparison

Just as the ratio of the frequency estimation RMSE of both methods are compared in the main text, we can also compare the amplitude estimates of the methods. Analogously to the definition in the main text, we define the ratio of amplitude RMSE as

$$\rho_{RMSE,A} = \log_2 \frac{\text{RMSE}_{A,EKS}}{\text{RMSE}_{A,SCF}}. \tag{6}$$

This ratio, alongside with slices through the parameter space, are displayed in Fig. 1. The ratio is below 0 everywhere, indicating that the amplitude time series estimation of the EKS is significantly better that the SCF estimation. This is due to the selection of the block length of the SCF: The block length that produces the lowest frequency RMSE is selected, which is often very long compared to the time scales of the exponential decay of the amplitude. Hence, the time resolution is too low to resolve the exponential decay, leading to large RMSE.

While the SCF block length could be optimized to produce a significantly lower amplitude RMSE, this would result in a larger frequency RMSE. This tradeoff can be avoided using the EKS, as the same model parameters produce both good frequency and amplitude estimates.

## 2.3. Validity of EKS uncertainty intervals

As the EKS is a Bayesian method, it inherently produces uncertainty estimates for its frequency estimates. However, since the EKS can only solve the nonlinear state space model approximately via linearization, we checked whether the uncertainty intervals accurately reflect the frequency estimation error.

The EKS assumes Gaussian probability distributions of its estimates. For each signal sample, the EKS produces a probability density function (PDF) that should

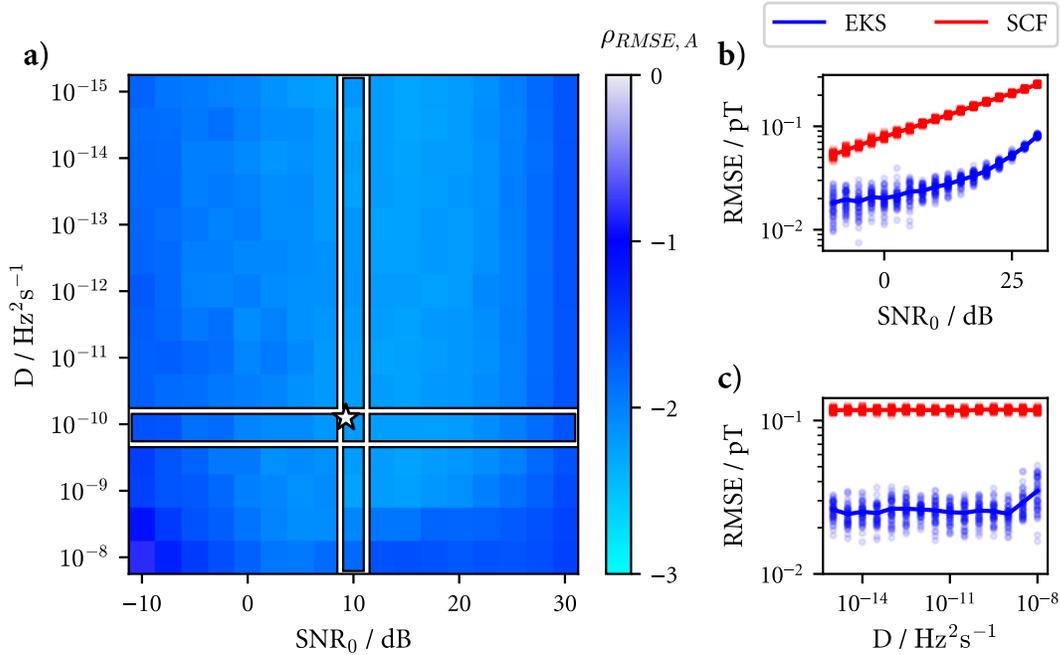

**Figure 1.** Quantitative comparison of the SinCos Fit against the EKS in a simulation study. **a)** $\log_2$ of the mean ratio of the amplitude estimate errors. Blue indicates a better performance of the EKS over the SCF, red the opposite. The star marker indicates a rough estimate of our experimental conditions. **b)** Horizontal slice through a). The RMSEs of individual simulation runs are displayed as markers, the solid lines indicate the mean. **c)** displays a vertical slice.

represent the probability of the value of the true frequency. Therefore, we can simply count the number of true frequency samples that lie within the $1\sigma$ interval of the estimated Gaussian PDF. If the estimated PDF represents the estimation error well, about 68% of true frequency samples of the full time series should fall within the $1\sigma$ interval.

This count is visualized in Fig. 2. In a large region of the parameter space, the count is very close to 68% of samples within the $1\sigma$ interval, indicating that the uncertainty estimates are valid. This regime corresponds to the region in which the EKS outperforms the SCF in frequency estimation. For very low frequency fluctuation and low SNR, the count is closer to 50%, which indicates that the uncertainty is underestimated. This coincides with the parameter regime where the SCF performs better than the EKS.

Not shown here is that the uncertainty estimate depends on the selected block length of the EKS. With small block lengths, the uncertainty tends to be underestimated while with large block lengths, it is overestimated. Therefore, the selection of the correct block length is critical if the uncertainty estimates are of importance for the application. Currently, no selection criterion other than comparison to simulations exists.

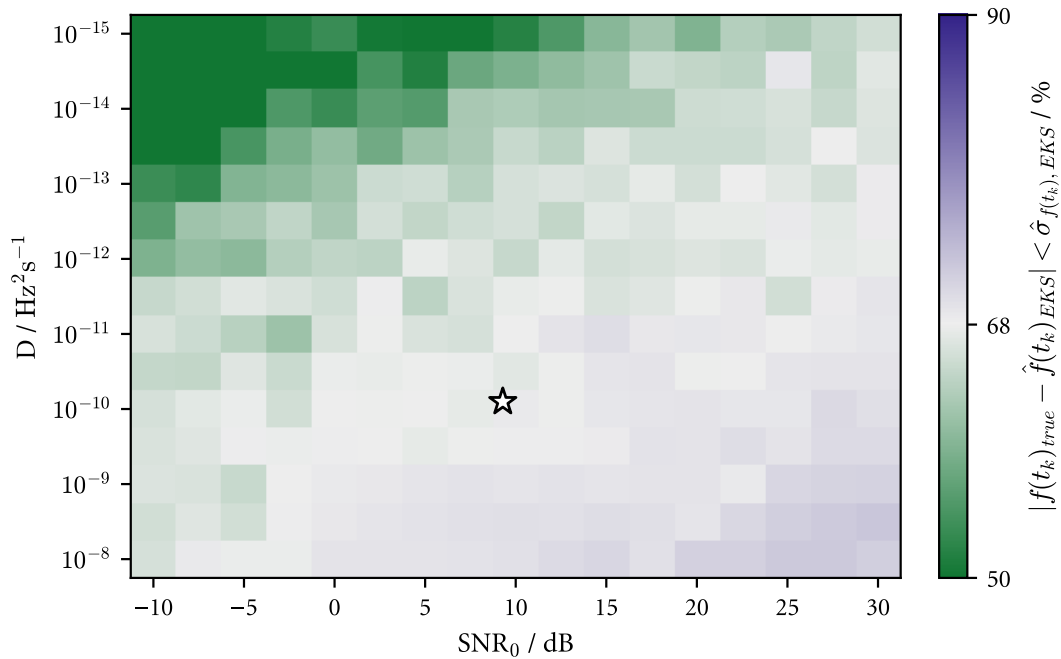

**Figure 2.** Percentage of true frequency samples $f(t_k)_{true}$ within the $1\sigma$ bounds of the EKS frequency estimate.

## 2.4. Optimal SCF block length

As mentioned in the main text, the optimal block length of the SCF is selected based on the minimal frequency estimation RMSE. In Fig. 3, these optimal block lengths are shown.

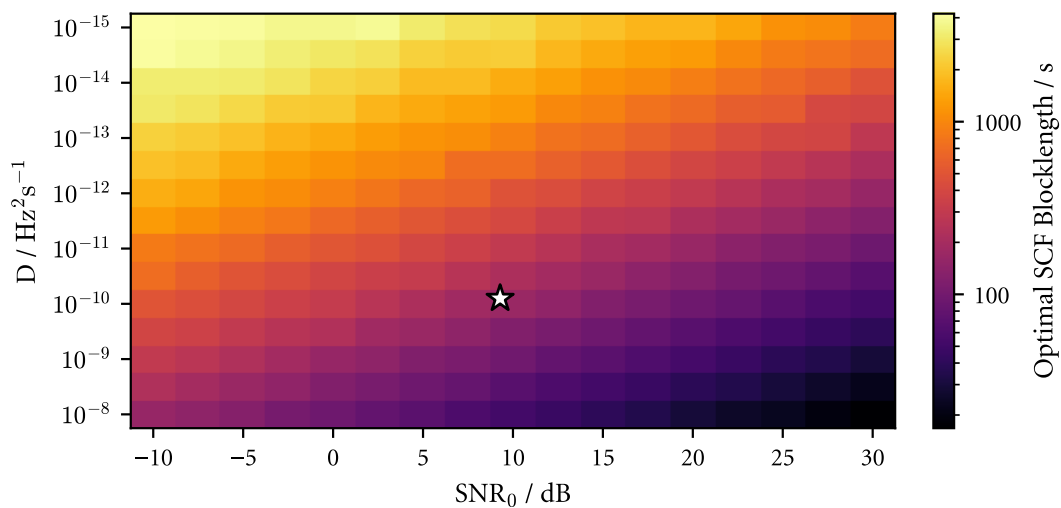

**Figure 3.** Optimal block length for SCF in the minimal frequency RMSE sense.

## 2.5. *Cramér–Rao lower bound (CRLB) for zero drift*

In the case of zero drift ($D = 0$) we can calculate the CRLB [3] and compare it to the performance of our estimators (see Fig. 4).

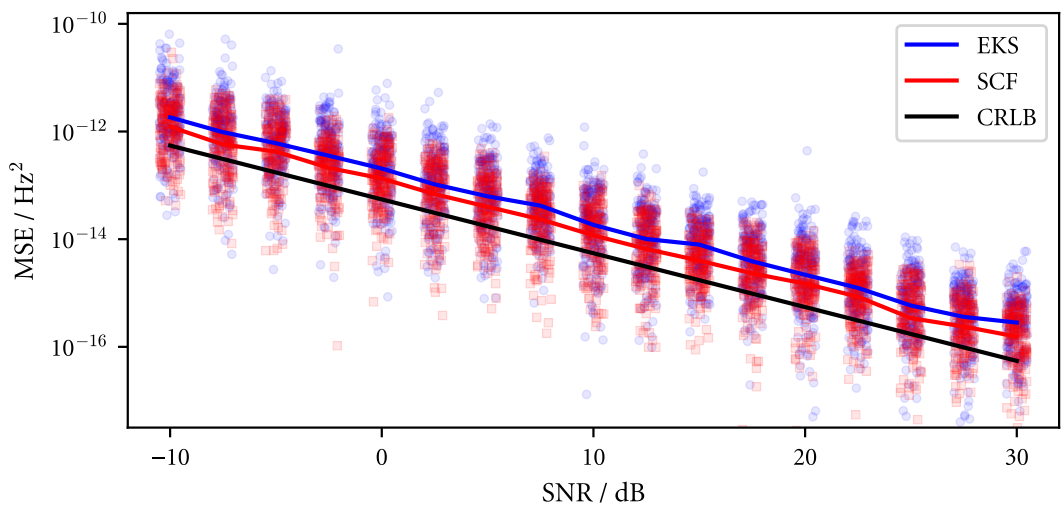

**Figure 4.** At zero drift the SCF performs better then the EKS, while both methods perform close to the theoretical variance limit of the CRLB.


\end{document}